\documentclass[%
reprint,
superscriptaddress,
 amsmath,amssymb,
 aps,
]{revtex4-2}

\usepackage{graphicx}
\usepackage{dcolumn}
\usepackage{bm}
\usepackage{amsmath}
\usepackage[dvipsnames]{xcolor}

\begin{document}

\preprint{APS/123-QED}

\title{Spin wave resonance in yttrium iron garnet stripe domains}

\author{Daniel Prestwood}
\email{daniel.prestwood.22@ucl.ac.uk}
\affiliation{London Centre for Nanotechnology, University College London,\\ London, WC1H 0AH, United Kingdom}
\affiliation{Department of Electronic and Electrical Engineering, University College London,\\ London, WC1E 7JE, United Kingdom}
\affiliation{Blackett Laboratory, Imperial College London, London SW7 2AZ, United Kingdom}

\author{Christopher E. A. Barker}
\affiliation{National Physical Laboratory, Hampton Rd, Teddington TW11 0LW, United Kingdom}

\author{Kilian D. Stenning}
\affiliation{Blackett Laboratory, Imperial College London, London SW7 2AZ, United Kingdom}

\author{Charlie W. F. Freeman}
\affiliation{London Centre for Nanotechnology, University College London,\\ London, WC1H 0AH, United Kingdom}
\affiliation{Department of Electronic and Electrical Engineering, University College London,\\ London, WC1E 7JE, United Kingdom}
\affiliation{National Physical Laboratory, Hampton Rd, Teddington TW11 0LW, United Kingdom}

\author{Tianyi Wei}
\affiliation{London Centre for Nanotechnology, University College London,\\ London, WC1H 0AH, United Kingdom}
\affiliation{Department of Electronic and Electrical Engineering, University College London,\\ London, WC1E 7JE, United Kingdom}

\author{Takashi Kikkawa}
\affiliation{Advanced Science Research Center, Japan Atomic Energy Agency, Tokai 319-1195, Japan}
\affiliation{Department of Applied Physics, The University of Tokyo, Tokyo 113-8656, Japan}

\author{Troy Dion}
\affiliation{Solid State Physics Laboratory, Kyushu University, Japan}

\author{Daniel Stoeffler}
\affiliation{Université de Strasbourg, CNRS, Institut de Physique et Chimie des Matériaux de Strasbourg,\\ UMR 7504, F-67000 Strasbourg, France}

\author{Yves Henry}
\affiliation{Université de Strasbourg, CNRS, Institut de Physique et Chimie des Matériaux de Strasbourg,\\ UMR 7504, F-67000 Strasbourg, France}

\author{Matthieu Bailleul}
\affiliation{Université de Strasbourg, CNRS, Institut de Physique et Chimie des Matériaux de Strasbourg,\\ UMR 7504, F-67000 Strasbourg, France}

\author{Noora Naushad}
\affiliation{NEST Research Group, Department of Computer Science, The University of Manchester, Oxford Road,\\ Manchester, M13 9PL, United Kingdom}

\author{William Griggs}
\affiliation{NEST Research Group, Department of Computer Science, The University of Manchester, Oxford Road,\\ Manchester, M13 9PL, United Kingdom}

\author{Thomas Thomson}
\affiliation{NEST Research Group, Department of Computer Science, The University of Manchester, Oxford Road,\\ Manchester, M13 9PL, United Kingdom}

\author{Murat Cubukcu}
\affiliation{London Centre for Nanotechnology, University College London,\\ London, WC1H 0AH, United Kingdom}
\affiliation{National Physical Laboratory, Hampton Rd, Teddington TW11 0LW, United Kingdom}

\author{Jack C. Gartside}
\affiliation{Blackett Laboratory, Imperial College London, London SW7 2AZ, United Kingdom}
\affiliation{London Centre for Nanotechnology, Imperial College London, \\London, SW7 2AZ, United Kingdom}

\author{Eiji Saitoh}
\affiliation{Department of Applied Physics, The University of Tokyo, Tokyo 113-8656, Japan}
\affiliation{Institute for AI and Beyond, The University of Tokyo, Tokyo 113-8656, Japan}
\affiliation{RIKEN Center for Emergent Matter Science (CEMS), Wako 351–0198, Japan}
\affiliation{WPI-Advanced Institute for Materials Research, Tohoku University, Sendai 980-8577, Japan}

\author{Will R. Branford}
\affiliation{Blackett Laboratory, Imperial College London, London SW7 2AZ, United Kingdom}
\affiliation{London Centre for Nanotechnology, Imperial College London, \\London, SW7 2AZ, United Kingdom}

\author{Hidekazu Kurebayashi}
\affiliation{London Centre for Nanotechnology, University College London,\\ London, WC1H 0AH, United Kingdom}
\affiliation{Department of Electronic and Electrical Engineering, University College London,\\ London, WC1E 7JE, United Kingdom}
\affiliation{WPI-Advanced Institute for Materials Research, Tohoku University, Sendai 980-8577, Japan}

\date{\today}

\begin{abstract}
We study a thin film yttrium iron garnet sample that exhibits magnetic stripe domains due to a small perpendicular magnetic anisotropy. Using wide-field magneto-optic Kerr effect measurements we reveal the domain pattern evolution as a function of applied field and discuss the role of the cubic anisotropy in the domain formation. Rich magnon spectra are observed in the stripe domain states, with a range of excitation conditions providing distinct spectra. The measurements are interpreted using micromagnetic simulations to provide the spatial profiles of each resonance mode. We further simulate domain patterns and resonance spectra accounting for the cubic anisotropy,with good correlation to experiment. This study highlights how non-collinear magnetic domain structures can host complex resonant behaviour in a low-damping magnetic material, with potential use in future magnonic applications. 

\end{abstract}

\maketitle


\section{Introduction}
Yttrium iron garnet (YIG) is a widely used material within the field of spintronics \cite{chumak_magnon_2015} and magnonics \cite{serga_yig_2010,Awschalom_IEEE2021,flebus_2024_2024} due to its low intrinsic damping (Gilbert damping constant $\alpha$ down to 5\(\times10^{-5}\)\cite{rao_liquid_2018}). For magnonic applications, YIG is typically studied in the collinear saturated magnetic state where the homogeneous ground state of magnetisation is effective in describing the spinwave excitations~\cite{damon_dispersion_1965,zhou_current-induced_2013,bhoi_study_2014,cornelissen_nonlocal_2017,satywali_microwave_2021,Kurebayashi_NMater2011,Makiuchi_NMater2024}. Non-uniform magnetic textures such as stripe, helical and vortex states, have been shown to display rich resonant modes with distinct spatially-distributed micro-spin characteristics. \cite{PhysRevLett.122.097202,PhysRevB.84.174401,gartside_reconfigurable_2022,lee_tunable_2021,hasty_Ferromagnetic}. Spinwaves in non-collinear magnetisation textures have been actively studied as a platform for unconventional computing applications e.g. reservoir computing \cite{nakane_spin_2021,lee_task-adaptive_2024,gartside_reconfigurable_2022,papp_nanoscale_2021,nakane_reservoir_2018}

Fabrication of YIG for magnonic devices often requires etching which may lead to surface damage and a reduction in material quality, including enhancing damping~\cite{heinz_propagation_2020}. Other techniques such as laser heating \cite{giacco_patterning_2024} and ion beam irradiation \cite{kiechle_spin-wave_2023} still present the inflexibility of making irreversible changes to the physical properties. On the other hand, magnetic domain structures naturally offer a route to reconfigure spin wave spectra and propagation without permanently altering material shape or properties~\cite{petti_review_2022,szulc_reconfigurable_2022}. For example, domain walls can act as nano-channels to guide spin waves \cite{wagner_magnetic_2016,garcia-sanchez_narrow_2015,henry_unidirectional_2019} or as local spin wave emitters~\cite{hollander_magnetic_2018,wang_spin_2014,van_de_wiele_tunable_2016}.

Micromagnetic domain patterns result from minimising local energetics, depending on factors including magnetic anisotropy, magnetic dipole interaction and applied field. High perpendicular magnetic anisotropy (PMA) materials often exhibit stripes, labyrinths and bubble domains \cite{salikhov_control_2022}, whereas materials with cubic anisotropy host Landau and fractal patterns in order to reduce the closure energy \cite{hubert_domain_1998}. Stripe domains are commonly found in thin films with small PMA \cite{fin_-plane_2015,tee_soh_magnetization_2013,camara_magnetization_2017,AitOukaci02092023}, often discussed with the condition \(Q = 2K_\perp/\mathrm{\mu_0} M_\mathrm{s}^2\)$<1$ with \(K_\perp\) being the uniaxial anisotropy constant and \(M_\mathrm{s}\) the saturation magnetisation~\cite{hubert_domain_1998}. These stripes can exist at zero external field when the film thickness is above the critical thickness, \(D_{\mathrm{cr}} = 2\pi\sqrt{A/K_{\perp}}\) where \(A\) is the exchange stiffness. For thicker films a Landau structure of flux closure domains can form at the film surface in order to reduce the surface energy \cite{hubert_domain_1998}.


Resonant properties of the stripe domain state have been studied in Bi and Ga
substituted YIG\cite{vukadinovic_nonlinear_2019,luhrmann_high-frequency_1993,ramesh_ferromagnetodynamics_1988} and others materials\cite{ebels_ferromagnetic_2001,vukadinovic_ferromagnetic_2001,camara_magnetization_2017,wei_top-down_2015,PhysRevB.96.024421,PhysRevB.105.094444}. These studies typically show multiple resonant modes in domains and domain-walls, and,  characteristic modes of flux closure caps that appear in thicker films \cite{ebels_ferromagnetic_2001}. Previous studies of the propagation of magnetostatic waves in YIG stripe domains~\cite{Vashkovskii1996,vashkovskii_propagation_1997,Vashkovskii1998} found that the different domain states could offer varying modes of propagation with a limited amount of theoretical and numerical analysis using a microspin model. 

In this study, we characterise both static and dynamic properties of 3 µm thick YIG thin films in a range of collinear and stripe domain states unde a small ($<$10mT) magnetic fields. For the dynamic properties, we concern ourselves primarily with the standing wave modes that are set up in the cross section of the stripe domain pattern. We observe how this regime enables mode hybridisation due to the confined nature and varying field-frequency trajectories of the various magnon modes.  Micromagnetic simulations provide spinwave spectra that agree well with our observations, allowing us to map out of the spatial distribution of the various resonant modes within different regions of the magnetic domain pattern. This study sheds light on the enhanced range of microstate and dynamic properties accessible in garnet films.

\section{Results}

\subsection{Stripe domain states and field evolution}
\begin{figure*}
\begin{centering}
\includegraphics[width=0.95\textwidth]{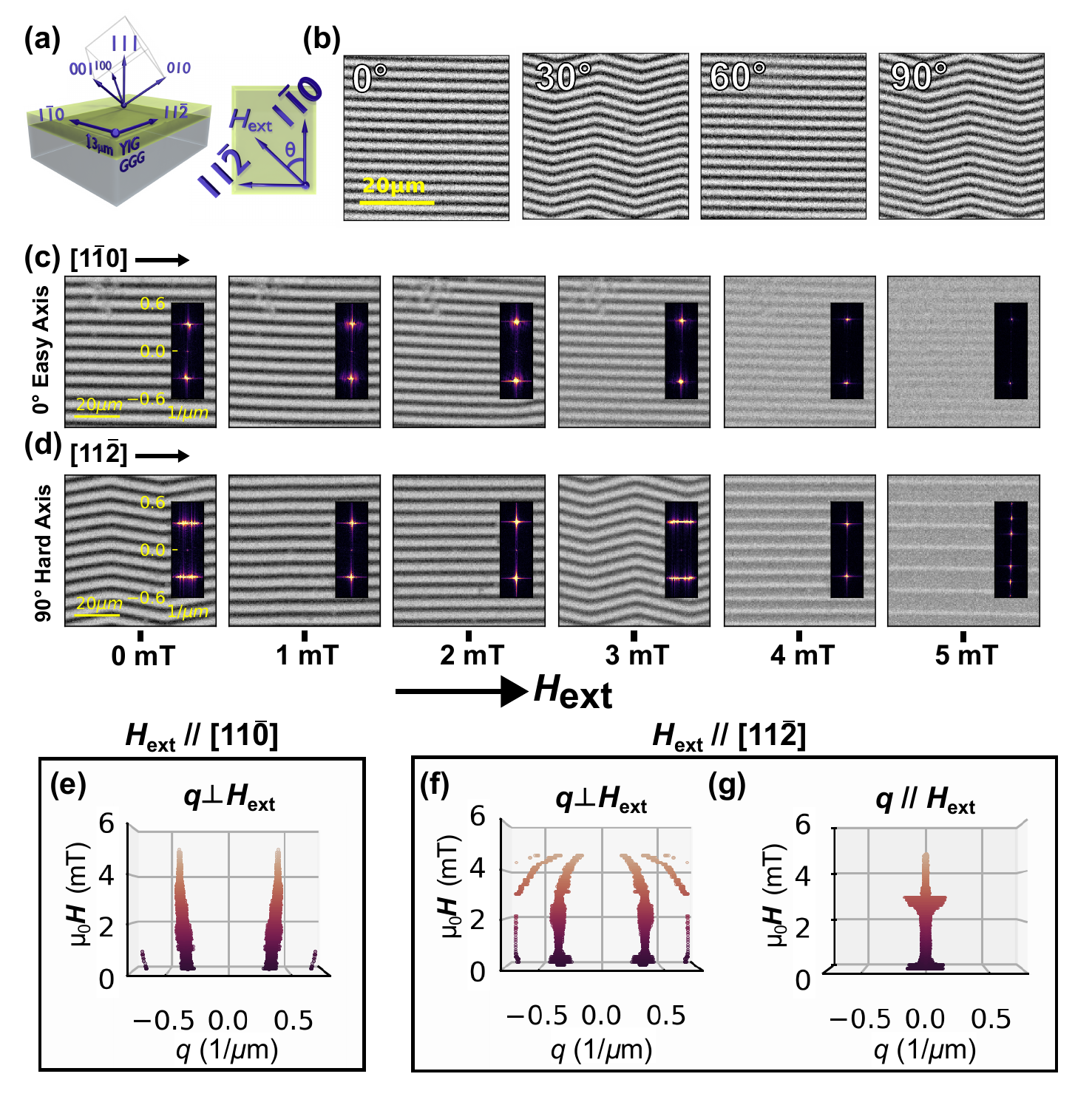}
\caption{(a) Diagram of the YIG sample with labelled crystalline axes. (b) Wide field Kerr images of magnetisation state at remanence after saturating \(\boldsymbol{H}_{\mathrm{ext}}\) of 17mT at 0, 30, 60 and 90$^\circ$ w.r.t the [1$\bar{1}$0] direction. c,d) Field evolution of the magnetic state for the field being applied along either [11$\bar{2}$] direction (c) or [1$\bar{1}$0] direction (d). Inset shows the Fourier transform of each image.  Position of Fourier transform peaks as a function of the \(\boldsymbol{H}_{\mathrm{ext}}\) for [1$\bar{1}$0] direction (e) and [11$\bar{2}$] direction (f-g). For [1$\bar{1}$0] direction, we see a smooth change from stripe to saturated consistent with a coherent rotation magnetisation process, whereas for the [11$\bar{2}$] direction, we see a transition into a state where one set of domains expand while the others shrink a process which is consistent with a domain propagation magnetisation process.}
\label{moke_crystal}
\end{centering}
\end{figure*}

The YIG crystal studied is a 3 µm thin film grown on a [111]-oriented gadolinium gallium garnet (GGG) substrate using liquid phase epitaxy (LPE)\cite{kikkawa_composition-tunable_2022,lee_nonlinear_2023} as illustrated in Fig.~\ref{moke_crystal}{(a)}. The PMA in the sample was determined by comparing in-plane and OOP hysteresis loops measured by a vibrating sample magnetometer and was found to be 1400 $\pm$ 200 Jm\(^{-3}\) (providing a \(Q\) value of 0.11). This PMA causes the formation of stripe domains as confirmed by magneto-optic Kerr effect (MOKE) images shown in Fig.~\ref{moke_crystal}{(b)}. Although the crystal structure of YIG does not allow any intrinsic uniaxial anisotropy, PMA in thin films can potentially arise from the inversion-breaking at the interface and tetragonal distortion due to growth-induced strain~\cite{coey_nanoscale_2010}. However, we rule out the interfacial contribution due to the thickness of our YIG sample. 
Growth-induced anisotropy has previously been observed in (111)-planed YIG in Refs. \cite{stacy_growthinduced_1972} and \cite{ya-qi_growth-induced_1986} where the GGG substrate was polished away to remove the contribution of stress from the substrate.

We begin by examining the remanent magnetisation state after saturating along various in-plane angles from the \(\langle1\bar{1}0\rangle\) direction which is determined from the substrate manufacturer specification (see Fig.~\ref{moke_crystal}{(a)}). As shown in Fig.~\ref{moke_crystal}{(b)}, 0\(^{\circ}\) and 60\(^{\circ}\) (directions referred to as effective easy axes), co-linear magnetic stripe domains are observed, whereas for 30\(^{\circ}\) and 90\(^{\circ}\) (directions referred to as effective hard axes), a zig-zag pattern emerges. As explained below, the six-fold symmetry of this effecive anisotropy in the (111) plane is consistent with the cubic anisotropy of YIG (\(\langle111\rangle\) easy axis). Stripe realignment also occurs for externally applied fields (\(\boldsymbol{H}_{\mathrm{ext}}\)) below saturation. As such, we ensure that a sufficiently large field was applied to align the stripe domains prior to measurement.

To further understand the evolution of stripe domains, we used MOKE microscopy to image the magnetisation state in an external field applied along the [1$\bar{1}$0] (effective easy) and [11$\bar{2}$] (effective hard) directions. The evolution of the stripes when the field is applied along the [1$\bar{1}$0] axis is shown in Fig.~\ref{moke_crystal}{(c)}. In this direction the stripe domains are collinear and aligned with the field. As the field strength increased, a gradual decrease of the stripe periodicity is observed from 2.8 $\pm$ 0.2 $\mu \mathrm{m}$ to 2.4 $\pm$ 0.1 $\mu \mathrm{m}$ prior to domain disappearance at \(\approx\) 5.5 mT - see the increase in wavenumber in the reciprocal mapping of Fig.~\ref{moke_crystal}{(e)}. In contrast, the domain evolution for fields applied along the [11$\bar{2}$] axis displays more significant changes as shown in Fig.\ref{moke_crystal}{(d)}. The stripes originate in the zig-zag state which quickly gives way to the collinear stripes at a relatively low field before gradually returning to the zig-zag state at 3 mT. After this, the domains straighten along the field axis. The periodicity of the stripe domains increases significantly from 2.8 $\pm$ 0.2 $\mu \mathrm{m}$ to 5.2 $\pm$ 0.1 $\mu$, with one set of stripes widening while the other narrows, before eventually saturating at \(\approx\) 5.5 mT as displayed more precisely in Fig.~\ref{moke_crystal}(f) which shows the reduction of the wavenumber along the [1$\bar{1}$0] direction. We can also identify the zig-zag state via the broadening in reciprocal space along the x-axis showing the finite periodicity in this direction. While there is a gradual transition from the collinear state to the zig-zag state, the transition from the zig-zag state to collinear at greater field is much more abrupt. Our observation is consistent with previous work for (111)-planed YIG~\cite{duda_domain_1974,vashkovskii_propagation_1997,Vashkovskii1999}, although the zig-zag state was not present in zero field. The effective easy and effective hard axis behaviour are representative of coherent rotation and domain propagation magnetisation processes, respectively. Thus far we have referred to the [11$\bar{2}$] and the[1$\bar{1}$0]directions as the effective hard and effective easy axes respectively as in reality the hard and easy axes are the \(\langle100\rangle\) and \(\langle111\rangle\) directions respectively. Further, given the crystalline direction of our sample, all the in-plane directions are equipotential, which we will show later on. However, the magnetisation in the sample in the small-field regime is not uniform, making the local magnetic calculations much more complex. In this article, we use the terms the \textit{effective} hard and easy axes for conciseness.


\begin{figure*}
\includegraphics[width=0.95\textwidth]{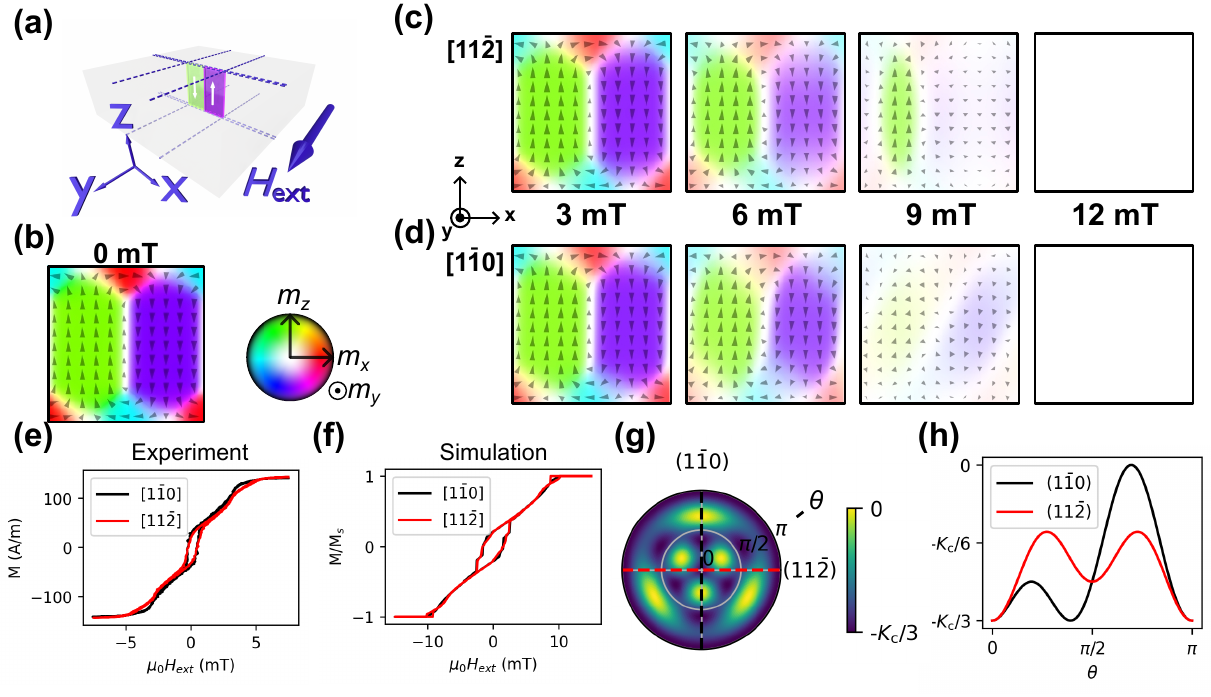}
\caption{Simulation of the remnant state of the simulation (a). With the micromagnetic state for increasing field along the [1$\bar{1}$0] (b) and [11$\bar{2}$] (c) axes. (d) Schematic showing the simulation geometry. Measured (e) and simulated (f) magnetisation loops for both the [1$\bar{1}$0] and [11$\bar{2}$] axes.}
\label{sim_cubic_static}
\end{figure*}

In order to explain the observed magnetic domain structures, we employ micromagnetic simulations (mumax3)~\cite{vansteenkiste_design_2014}. Figure~\ref{sim_cubic_static}{(a)} shows a vertical section of the thin film with the height of the simulation set to the film thickness and the width to the stripe period. To do this we use a cell size of (16.67nm $\times$ 16.67nm $\times$ 16.67nm) with a grid size of (180 $\times$ 180 $\times$ 1). Periodic boundary conditions were then used to ensure a realistic demagnetising field.
As shown in Fig.~\ref{sim_cubic_static}b, stripe domains are successfully reproduced as the equilibrium state after removing the field, consistent with our MOKE observation. Flux closure domains are present at the top and bottom surfaces, characteristic of thicker films \cite{kittel_theory_1946,hubert_domain_1998}. Results by applying a field along the [1$\bar{1}$0] and [11$\bar{2}$] directions agree well with the MOKE images in Fig.~\ref{moke_crystal}. In these simulations \(\boldsymbol{H}_{\mathrm{ext}}\) is applied along the stripe direction (which is perpendicular to the simulated cross-section). It should be noted that we only simulate the cross section of the stripe domains. Not only does this limit the wave vector directions of spin waves that we can observe to a two-dimensional plane, it also prevents the possibility of simulating the zig-zag state that we observe when applying a field along the hard axis. As we will show in this article, much of the experimental results are captured by these simulations.
    
Figure~\ref{sim_cubic_static}{(c)} shows that the increasing periodicity of the stripes by the in-plane field is reproduced in the [11$\bar{2}$] simulation. For the [1$\bar{1}$0] case, we are able to observe a behaviour not elucidated by MOKE, i.e. a slanting of the stripe domains under the \(\boldsymbol{H}_{\mathrm{ext}}\) (Fig.~\ref{sim_cubic_static}{(d)}).
Figures~\ref{sim_cubic_static}{(e-f)} show the measured and simulated magnetisation loops for the two different axes. While the general shape of these matches well with a small hysteresis in the centre of the loop, one visible disagreement is that the width of hystereses in the simulation is larger than those in experiments, as is the saturation field. We speculate that is due to the fixed lateral period being imposed on the sample rather than being optimized for every field value in addition to effects from temperature reducing the coercive field for the experiment compared to simulation - further discussions can be found in Appendix \ref{Sim Params}. The qualitative agreement lead us to use these simulation parameters in the present study.

The field evolution of the domain pattern can be explained by examining the first-order cubic anisotropy energy density, \(E\),(Fig.~\ref{sim_cubic_static}{(g)}) given by the following \cite{coey_ferromagnetism_2010}:

\begin{multline*}
    E = K_{\mathrm{c}} \left( \frac{\sin^4\theta}{4} +  \sqrt{2}\cos^2\phi \cos\theta\sin\phi\sin^3\theta \right. \\  + \left. \frac{\cos^4\theta}{3} + \frac{\sqrt{2}\cos\theta\sin^3\phi\sin^3\theta}{3} \right)
\end{multline*}
where \(\theta\) is the polar angle from the [111] direction and\(\phi\) is the azimuthal angle in the (111) plane such that \(\phi = 0\) points along the [1$\bar{1}$0] direction. The three hard axes directed along the \(\langle\)100\(\rangle\) directions. Using this we can examine three important case, the energy density for the (111),(1$\bar{1}$0) and (11$\bar{2}$) planes. The anisotropy energy in (111) plane, $E_{(111)}$, (i.e. the plane that defines the film and contains the [11$\bar{2}$] and [1$\bar{1}$0] directions) is an equipotential within the plane expressed as \(E_{(111)} = K_{\mathrm{c}}/4\) as we previously mentioned. As such, there is no hard or easy axis within the plane. 

We can also calculate the energy density as a function of polar angle from the [111] direction in the (11$\bar{2}$) (containing the [1$\bar{1}$0] direction) and (1$\bar{1}$0) (containing the [11$\bar{2}$] direction) planes, $E_{(11\bar{2})}$ and $E_{(1\bar{1}0)}$ respectively, as follows:
\begin{equation}
\label{e_perp_112}
    E_{(11\bar{2})} = K_\mathrm{{c}} \left( \frac{\cos^4\theta}{3} + \frac{\sin^4\theta}{4}\right)
\end{equation}
\begin{equation}
\label{e_perp_110}
    E_{(1\bar{1}0)} = K_{\mathrm{c}} \left( \frac{\cos^4\theta}{3} + \frac{\sqrt{2}\cos\theta\sin^3\theta}{9} +\frac{\sin^4\theta}{4}\right)
\end{equation}
From this we can observe that \(E_{(1\bar{1}0)}\) has a term that is odd in \(\theta\) around the in-plane direction, \(\theta=\pi/2\)  (Fig.~\ref{sim_cubic_static}{(h)}). As such, when \(\boldsymbol{H}_{\mathrm{ext}}\) is applied along the [11$\bar{2}$] (effective hard axis) the two stripes begin to cant towards film plane. As the in plane component of the magnetisation increases, one stripe stipe domain will approach a maxima of the anisotropy energy while the domain with the opposite OOP component of the magnetization will approach an energy minima. As such one set of stripes will expand while the other shrinks as one stripe magnetisation becomes more favourable than the other. Conversely, when magnetising along the [1$\bar{1}$0] (easy axis) direction we can see that \(E_{(11\bar{2})}\) is completely even in \(\theta\) (Fig.~\ref{sim_cubic_static}{(h)}) and as such neither stripe is favoured. By more closely examining the energy density (Fig.~\ref{sim_cubic_static}{(g)}) we can also see that as the stripe is magnetising into the film plane it becomes energetically favourable for the magnetisation of the stripes to tilt slightly explaining the behaviour close to saturation seen in the simulations.

\subsection{Stripe domain resonances}

We performed rf absorption experiments by placing the YIG sample on a broadband waveguide. The rf transmission coefficient $S_{12}$ as a function of frequency and applied magnetic field using a vector network analyser. We apply external dc magnetic field either perpendicular or parallel to the oscillating magnetic field, so-called perpendicular and parallel pumping geometry, to capture all the modes excited in the unsaturated regime. We repeat the same experiments after rotating the sample by 90 degrees in-plane. While in simulation the stripes are aligned along the the field direction by design, in the experiment we apply an initial saturating field prior to each scan to ensure that the stripes are aligned along the field direction. It should be noted that since the sample is laterally larger than the signal line, an out of plane (OOP) component of the rf excitation for both pumping regimes is present in the YIG sample. Furthermore, because the large width of the signal line (0.5 mm) we assume a uniform excitation given the comparatively smaller domain width.
 
\begin{figure*}
\includegraphics[width=0.95\textwidth]{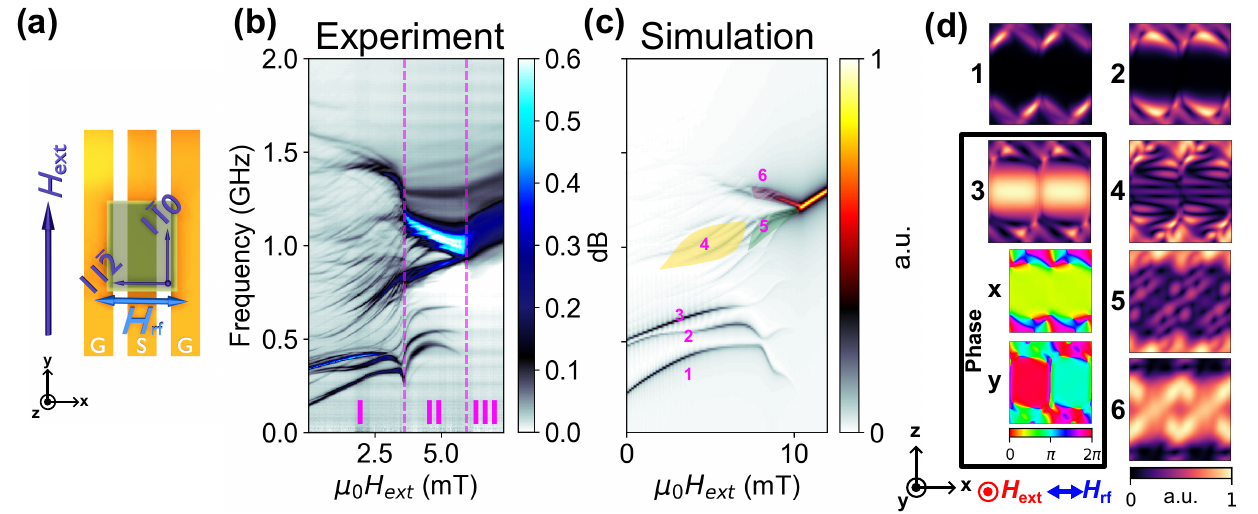}
\caption{a) Diagram showing the relative directions of external field, rf field and anisotropy axes. (b,c) Measured (b) and simulated (c) magnetic resonance spectra for YIG sample for rf \(\perp\) $\boldsymbol{H}_{\mathrm{ext}}$ which is directed along the [1$\bar{1}$0] axis. (d) power maps of spin wave modes labelled on (b) with a phase map shown for the bulk mode (3). The $x$ and $z$ axis are directed horizontally and out of the page respectively. With the $z$ direction being along the axis of the stripe domains and the magnetisation within the stripes pointing along the $y$ axis at zero field. These power maps are taken at 3mT (1-4) and 9.5mT (5 and 6).}
\label{fmr_easy_perp}
\end{figure*}

Figure~\ref{fmr_easy_perp} presents spinwave spectra and simulation results for the perpendicular pumping with \(\boldsymbol{H}_{\mathrm{ext}}\) applied along the easy axis Fig.~\ref{fmr_easy_perp}(a). We divide the spectra in Fig.~\ref{fmr_easy_perp}(b) into three distinct field regimes. A low-field regime for 0 - 3.5 mT (I) has clear stripe domain mainly polarised along the OOP directions; there is an intermediate field regime from 3.5 - 6 mT (II) which is associated with a slanting of the stripe domains towards the nearest easy axis that only partially points OOP, according to our simulations; finally the moments are collinear to the applied field in the high-field regime (III) above 6 mT where the Kittel mode appears. Using the same parameters for the prior static simulations, we calculated the resonant spectra as shown in Fig. \ref{fmr_easy_perp}(c). Three modes labelled as 1-3 on Fig.~\ref{fmr_easy_perp}(c) are identifiable in the experimental spectra in a very similar frequency region. Their spatial power maps (Fig.~\ref{fmr_easy_perp}(d)), generated by plotting the Fourier transform amplitude for each cell at a given frequency, reveal that the lowest-frequency mode (Mode 1) is the domain wall mode associated with the 90$^{\circ}$ domain walls between the stripe domains and flux closure caps. Mode 2 has strong resonance intensity at the top and bottom surfaces of the bulk stripe domains which have a tilted magnetisation from the bulk moments as determined from the static simulations. Mode 3 is a bulk mode of the stripe domains themselves. The phase maps shown for the bulk mode (3) reveal the two stripe domains to be in phase along the $x$-axis but out of phase along the $y$-axis which is the axis along which the rf field is applied. This phase relationship is indicative an acoustic mode as we would expect for the direction of the rf field being applied here and is the case for all the modes excited under this rf direction.

Modes resonating at a higher frequency than Mode 1-3 show a more hybridised character. This is apparent in the spatial mapping of the example mode for Region 4 in Fig.~\ref{fmr_easy_perp}(d), where high-order bulk modes resonate together with modes located in the flux closure domains. The modes in this region are characterised by a finite wavevector parallel to the static magnetisation within the stripes, determined through Fourier analysis of the spatial power maps. The modes in Regions 4-6 are much more complex in terms of hybridisation where analysing individual modes does not offer much insight into the dynamics. The two modes from Regions 5 and 6 that meet at the saturation field (\(\boldsymbol{H}_{\mathrm{ext}}\) $\approx$ 10 mT) can be also be identified in the experiment. The power maps reveal these modes in these regions are the surface (5) and bulk (6) modes.

\begin{figure*}
\includegraphics[width=0.95\textwidth]{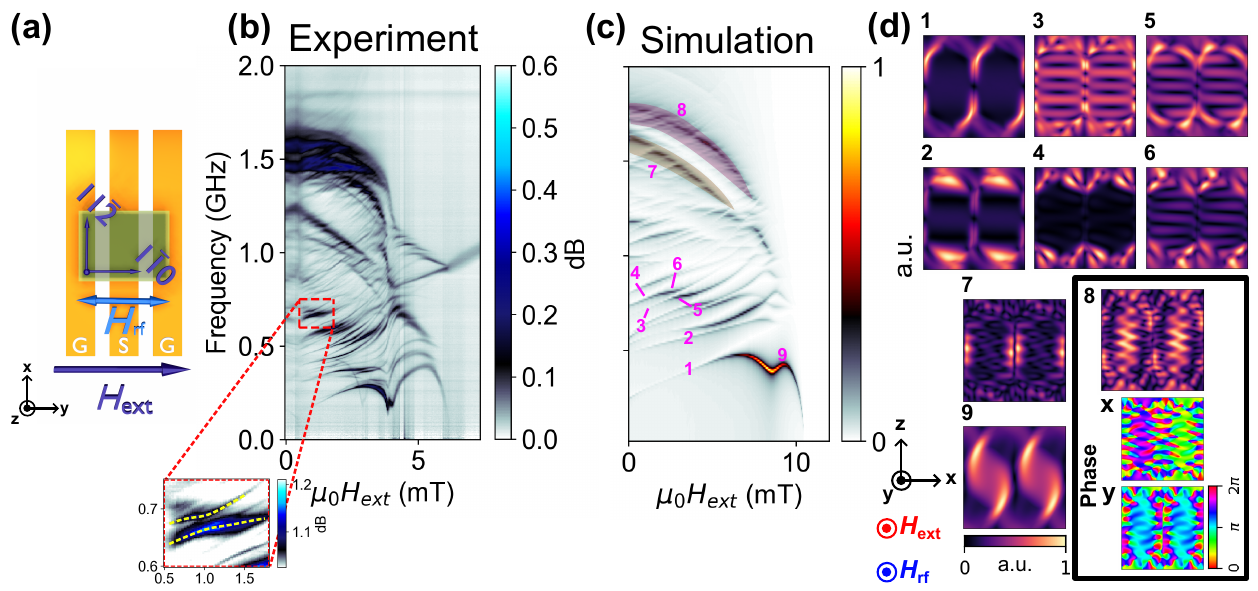}
\caption{a) Diagram showing the relative directions of external field, rf field and anisotropy axes. (b,c) Measured (b) and simulated (c) magnetic resonance spectra for YIG sample for rf \(\parallel\) $\boldsymbol{H}_{\mathrm{ext}}$ which is directed along the [1$\bar{1}$0] axis.  (d) power maps of spin wave modes labelled on (c) with a phase map shown for the bulk mode (6). These power maps are taken at 3mT (1,2,5 and 6), 1mT (3,4,7 and 8) and 8mT (9).}
\label{fmr_easy_par}
\end{figure*}

Figure \ref{fmr_easy_par} shows spinwave spectra for the parallel pumping configuration and corresponding micromagnetic simulation results for \(\boldsymbol{H}_{\mathrm{ext}}\) along the easy axis. Here we can also identify Modes 1 and 2 which are localised in the same regions as Modes 1 and 2 from the perpendicular pumping results. The difference in field frequency profile is likely explained by the rf field direction which determines the phase relationship of the dynamic component of the magnetisation between the stripes (Fig.~\ref{fmr_easy_par}{(d)}). The the power maps for these modes also indicates a difference in the hybridisations that occur.  Interestingly, these modes primarily associated with the domain walls are not excited efficiently at lower fields for parallel pumping. This can be explained by the fact that magnetisation in these regions is predominantly parallel to the rf field applied. This does not change as larger fields are applied, suggesting that the hybridisation with other modes allow these regions to be more efficiently excited. Modes 3-6 in Fig.~\ref{fmr_easy_par}{(d)} exemplify the presence of hybridisation in our sample. This particular coupling can be observed both in the simulation (as denoted by the numbers 3-6) in addition to the experiment experiment (red dotted box). Modes 3 and 4 in Fig.~\ref{fmr_easy_par}{(d)} show the spatial profiles of the modes away from the anti-crossing (\(\boldsymbol{H}_{\mathrm{ext}}\) $\approx$ 1 mT) which reveal these two modes to be an edge mode and the 5th-order thickness modes. For the spatial profile for Modes 5 and 6 either side of the anti-crossing (\(\boldsymbol{H}_{\mathrm{ext}}\) $\approx$ 3 mT), we can observe that the two modes have hybridised, both showing aspects of the spatial profile present in Modes 3 and 4.  For the higher frequency modes, we encounter a similar issue as mentioned earlier, whereby the prominent role of hybridisation leads us to discuss the modes in the shaded regions (7 \& 8) by their common behaviour. Modes in Region 7 display resonance mainly in the 90$^{\circ}$ domain walls, whereas those in Region 8 are associated with bulk modes having a two-dimensional standing-wave nature along the parallel and perpendicular to the thickness direction ($z$-coordinate in Fig.\ref{fmr_easy_par}), coupled to the edge modes. 

By following Mode 1 to the high field regime where we see a discontinuous change in the resonant modes, which we label as Mode 9, we can attempt to understand the intermediate phase in our experiment. As we discussed earlier using Fig.~\ref{sim_cubic_static}(d), the stripe domains pick up a magnetisation in-plane perpendicular to the applied field, as the cross section of the stripes also slants in the same direction. Evident from the the power map of Mode 9 in Fig~\ref{fmr_easy_par}, the slanting rotates the direction of the $k$-vector of these spin waves, increasing the wavelength of the first-order mode in tandem with any other higher-order modes.

\begin{figure*}
\includegraphics[width=0.95\textwidth]{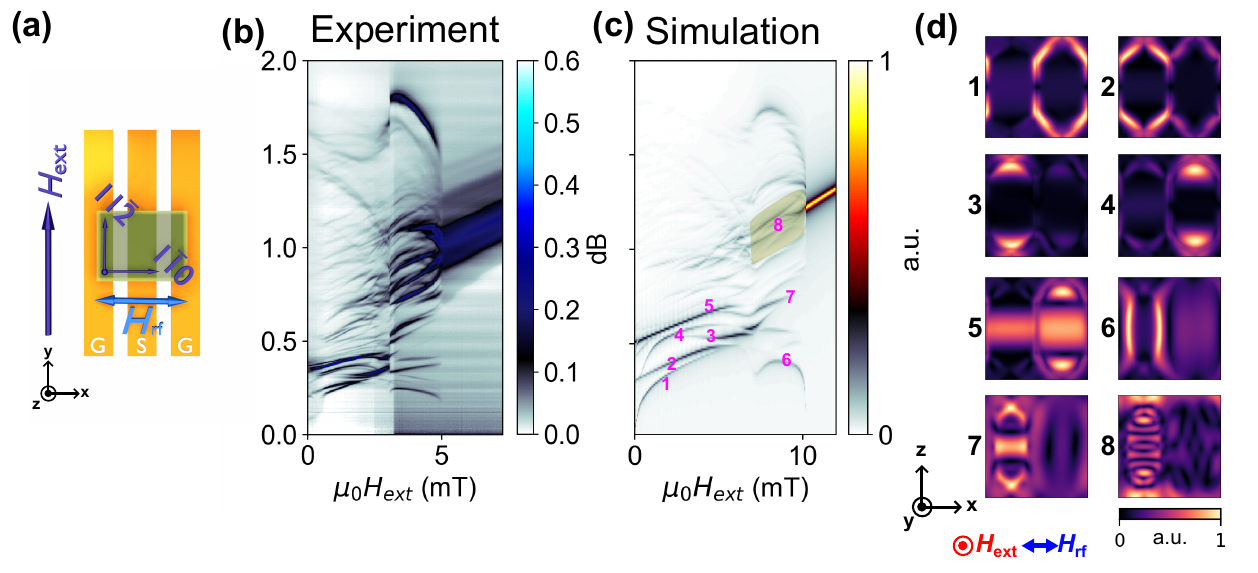}
\caption{a) Diagram showing the relative directions of external field, rf field and anisotropy axes. (b,c) Measured (b) and simulated (c) magnetic resonance spectra for YIG sample for rf \(\perp\) $\boldsymbol{H}_{\mathrm{ext}}$ which is directed along the [11$\bar{2}$] axis.  (d) power maps of spin wave modes labelled on (c). The power maps were taken at 3mT (1-5) and 8mT (6-8).}
\label{fmr_hard_perp}
\end{figure*}

We now turn to the case where \(\boldsymbol{H}_{\mathrm{ext}}\) is applied along the hard axis, which is summarised in Fig.~\ref{fmr_hard_perp}, showing a difference in spectra to Fig.~\ref{fmr_easy_perp}, particularly in the intermediate-field region. We attribute this to the evolution of static magnetic states, i.e. for the easy axis we observed a slanting of the stripes away from the perpendicular direction as we increased magnetic field, while for the hard axis an expansion of one set of stripes along side the shrinking of the other occurs. This ground-state difference impacts on the resonant spectra for the intermediate-field regime (3.5 mT - 5 mT in experiment and 7 mT - 10 mT in simulation), together with noticeable difference in the low-field regime below 3.5 mT. The most notable feature is the energy splitting of the two stripes for the low frequency modes (a splitting of Modes 1-2 and Modes 3-4 in Fig.~\ref{fmr_hard_perp}(d)) which share the same relative localisation within each stripe. The power maps in Fig.~\ref{fmr_hard_perp}(d) reveal Modes 1 and 2 to be domain wall modes but relating to different stripes magnetisation. Mode 2, active in domain walls of stripe magnetised in the positive OOP direction, has slightly higher frequency than Mode 1, which is associated with the stripe magnetised in the negative OOP direction. The frequency splitting of these modes is explained via the same mechanism that causes differing widths of the two stipes at higher fields (Fig.\ref{sim_cubic_static}(c)). The cubic anisotropy creates a symmetry breaking regarding the mirror symmetry with respect to the mid-plane fo the film, as demonstrated in Figs.~\ref{sim_cubic_static}(g-h), provided that there is some tilt towards the [11$\bar{2}$] direction. The asymmetry of the cubic anisotropy leads to both a slight difference in the magnetisation of each stripe as well as a difference in the anisotropy energy density.  Modes 6-7 and those in Region 8 reveal how the intermediate state where we observe uneven widths of the stripe domains can affect the resonant spectra primarily by altering the fundamental wavelength of the magnons along the the $x$-direction. Mode 6, like Modes 1 and 2, is predominantly excited in the domain wall. However, in contrast to Modes 1 and 2, this is localised in the domain-wall region where the magnetisation rotation is close to a 90\(^{\circ}\) difference. The complex configuration of dipolar fields in this state makes specific analysis of the modes non-trivial, but certainly the change in domain width in this region alters the wavelength of the standing wave modes with a \(k\)-vector directed across the domains.   
We see the reverse for Modes 3 and 4 where the magnetised negatively along the $z$-direction has a higher energy for this mode localisation. 

\begin{figure*}
\includegraphics[width=0.95\textwidth]{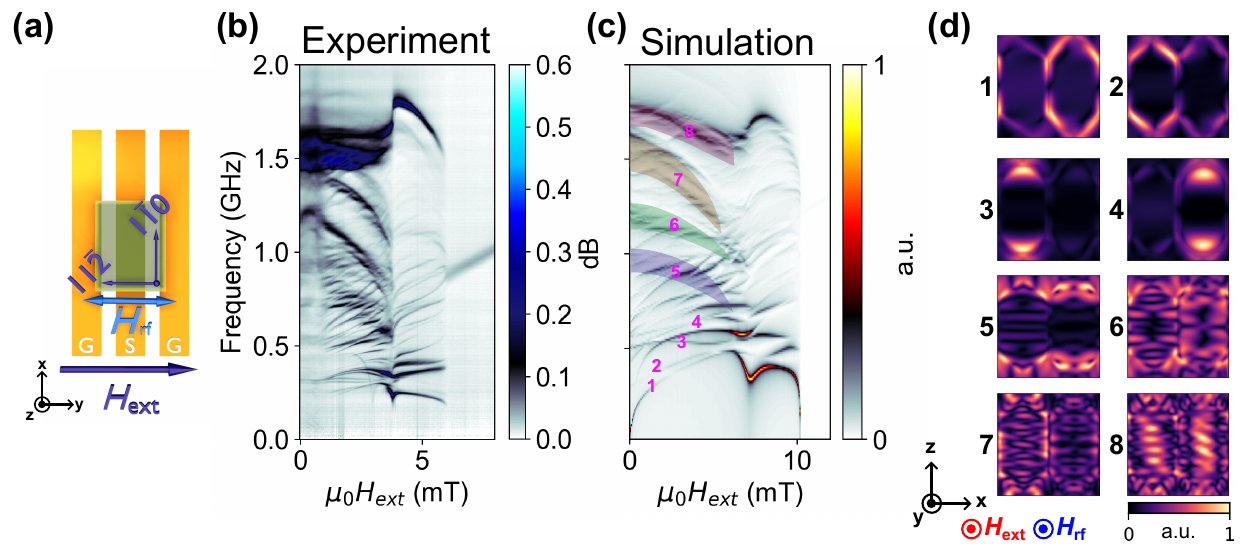}
\caption{a) Diagram showing the relative directions of external field, rf field and anisotropy axes. (b,c) Measured (b) and simulated (c) magnetic resonance spectra for YIG sample for rf \(\parallel\) $\boldsymbol{H}_{\mathrm{ext}}$ which is directed along the [11$\bar{2}$] axis. (d) power maps of spin wave modes labelled on (c). The power maps were taken at 3mT (1-4), 1mT (5,7 and 8) and 2mT (6).}
\label{fmr_hard_par}
\end{figure*}

Finally, we show the last experimental geometry for the parallel pumping with \(\boldsymbol{H}_{\mathrm{ext}}\) along the hard axis in Fig.~\ref{fmr_hard_par}. The same energy splitting of the lower frequency modes is present and Regions 5-8 appear to follow the same trend as those in Fig.~\ref{fmr_easy_par}(c). A substantial change in the resonant spectra in the intermediate state is observed as in Fig.~\ref{fmr_hard_perp}, though the relative excitation efficiency of positive and negative gradient modes is reversed between the two. This is likely due to the collinearity of magnetisation with the external field, as we would expect parallel pumping to be inefficient.

Taken together, the trend observed from the four experiments is that the resonant modes which couple to the excitation in the stripe domain phase are primarily determined by the pumping configuration. Most notably we selectively excite acoustic and optical modes for perpendicular and parallel rf excitation respectively. Mode hybridisation across the different edge and bulk sections is observed due to local exchange and dipole interactions. The good qualitative agreement of the simulated resonant spectra with the experimental ones demonstrate that, despite the discrepancy of the saturation field, the simulation parameters chosen still allow us to elucidate the origin of the resonant mode observed. At higher fields magnetic anisotropy determines the domain evolution in the ground state, on which different resonant modes are excited. This offers a potential opportunity to enhance the reconfigurability of magnon systems by manipulation of the domain state.

\section{Conclusion and outlook}
\label{conclusion}

We have studied the resonant spectra of the stripe domain states in 3 µm thick LPE-grown YIG thin films. Due to its low damping, we resolved multiple spinwave modes, where micro-magnetic simulations are a powerful tool to understand the spatial distribution of both static and dynamic components of magnetisation, e.g. modes populated in the flux closure caps and domain walls. We have shown how the small cubic anisotropy is able to dramatically affect the domain structure as a function of applied magnetic field and therefore the nature of resonant modes on the background magnetisation. The PMA present in our sample is not intrinsic to the crystal structure and as such may be easily controlled by growth conditions e.g. growth direction in the case of growth-induced anisotropy \cite{stacy_growthinduced_1972,ya-qi_growth-induced_1986}. Application of strain \cite{wang_strain-tunable_2014,gross_voltage_2022} or even spin currents \cite{chen_signatures_2024} leads to potential opportunities for tuning this relatively weak anisotropy, realising reprogrammable magnonic modes excited on the stripe domain states. Observed resonances depending on the pumping geometry allow for another level of selectivity, controlled by small magnetic fields (1 - 4 mT) as demonstrated. This differentiates from other methods such as near infra-red pulses \cite{gidding_reorientation_2024}. The stripe direction being influenced by the cubic anisotropy at zero field opens up another pathway to reorient these stripes and as such to change the excited resonances~\cite{yu_magnetic_2021}. Our study is highly relevant to a recent study of magnetic stripes in La$_{0.67}$Sr$_{0.33}$MnO$_{3}$, exhibiting similar resonant and propagation modes tunable via stripe direction~\cite{zhang_switchable_2025}. The ability to switch from a stripe domain state to completely saturated state via the application of a relatively small field created an opportunity for multifunction devices that make use of both states for different purposes e.g. data processing and transfer.

\begin{acknowledgments}
D.P. is supported by the EPSRC and SFI Centre for Doctoral Training in Advanced Characterisation of Materials Grant Ref: EP/S023259/1

This project was partially supported by the UK Government Department for Science, Innovation and Technology through National Measurement System funding (Metrology of complex systems for low energy computation).

T.K. and E.S. acknowledge support from JST CREST (JPMJCR20C1 and JPMJCR20T2), Grant-in-Aid for Scientific Research (Grants No. JP19H05600 and JP24K01326), and Grant-in-Aid for Transformative Research Areas (Grant No. JP22H05114) from JSPS KAKENHI, MEXT Initiative to Establish Next-generation Novel Integrated Circuits Centers (X-NICS) (Grant No. JPJ011438), Japan, and the Institute for AI and Beyond of the University of Tokyo.

M.B., D.S. and Y.H. thank Paul Noël for instructive discussions and acknowledge financial support by the Interdisciplinary Thematic Institute QMat, as part of the ITI 2021-2028 Program of the University of Strasbourg, CNRS and Inserm, IdEx Unistra (ANR 10 IDEX 0002), SFRI STRAT’US Project (ANR 20 SFRI 0012), and ANR-17-EURE-0024 under the framework of the French Investments for the Future Program

This work was supported by the Royal Academy of Engineering Research Fellowships, awarded to JCG.
JCG was supported by the EPSRC ECR International Collaboration Grant EP/Y003276/1.

\end{acknowledgments}

\appendix

\section{Simulation Parameters}
\label{Sim Params}

We use mumax3 \cite{vansteenkiste_design_2014} to simulate the experimental results. To do this, we simulate a vertical section of the thin film with the height of the simulation set to the film thickness and the width to the stripe period. Periodic boundary conditions were then used to simulate the demagnetizing field. A schematic of the geometry is shown in Fig \ref{sim_cubic_static}{\bf d}. For the simulations, we used the anisotropy constants of K$_\text{u}$ = 1000 Jm$^{-3}$ and K$_\text{c}$ = -600 Jm$^{-3}$. The K$_c$ value is an approximate value from literature \cite{lee_ferromagnetic_2016,hansen_anisotropy_1974}, while the K$_\text{u}$ value is chosen to best reproduce the experimental results. This value somewhat lower than our measured value, and yet, the saturation field we see in simulation is approximately double that of the experiment. As the predominant source of anisotropy in our sample is the demagnetizing field. Our method for calculating the uniaxial anisotropy involves subtracting the total anisotropy field from the demagnetizing field that we would expect for a thin film with the given measured value of M$_s$. Therefore, any error in the value of M$_s$ has a much more significant effect on the calculated K$_\text{u}$. As an example, a 2\% change in the M$_s$ value would lead to an over 35\% change in K$_\text{u}$.

We also use an exchange stiffness value of 6.5\(\times\)10\(^{-12}\)J/m a value that is somewhat larger than the commonly reported value of \(\approx\) 4\(\times\)10\(^{-12}\) J/m \cite{klingler_measurements_2014,slonczewski_temperature_1974,matsumoto_optical_2018}. While for thin films the stripe domain width is typically equal to the film thickness, \(D\), for thicker films the optimal width takes the form of
\begin{equation}
    W_\text{opt} = 2\sqrt{2D\sqrt{A/K_{\perp}}}.
\label{wopt}
\end{equation}
Using (\ref{wopt}) to calculate the optimal stripe domain width for the lower value gives \(W_\text{opt} = 1.23 \mu m\), we then observe that even when the simulation is initiated with stripe domains of period \(1.5\mu m\) as in the experiment, under applied field these domains will divide to produce a shorter domain width. While a value of 9\(\times\)10\(^{-12}\) J/m is required for \(W_\text{opt} = 1.5 \mu m\), we choose 6.5\(\times\)10\(^{-12}\) J/m as a balance between these two options such that we do not see the division of the domains under applied field.




\bibliography{references}

\end{document}